\documentclass[aps,amssymb,showkeys]{revtex4}

\usepackage{amsmath}
\usepackage{graphicx}
 \def\p{\partial}

\usepackage[dvips, bookmarks, colorlinks=true, plainpages = false, citecolor = red, urlcolor = blue, filecolor = blue]{hyperref}

\newcommand{\be}{\begin{equation}}
\newcommand{\ee}{\end{equation}}
\newcommand{\bea}{\begin{eqnarray}}
\newcommand{\eea}{\end{eqnarray}}
\newcommand{\ba}[1]{\begin{array}{#1}}
\newcommand{\ea}{\end{array}}
\renewcommand{\l}{\lambda}
\renewcommand{\a}{\alpha}
\renewcommand{\b}{\beta}

\begin{document}
\title{Percolation Crossing Formulas and Conformal Field Theory}
\author{Jacob J. H. Simmons}
\email{Jacob.Simmons@umit.maine.edu} 
\affiliation{LASST and Department of Physics \& Astronomy,
University of Maine, Orono, ME 04469, USA}
\author{Peter Kleban}
\email{kleban@maine.edu}
\affiliation{LASST and Department of Physics \& Astronomy,
University of Maine, Orono, ME 04469, USA}
\author{Robert M. Ziff}
\affiliation{Michigan Center for Theoretical Physics and Department of Chemical Engineering, University of Michigan, Ann Arbor MI 48109-2136}
\email{rziff@umich.edu}

\date{\today}
\begin{abstract} 
 Using conformal field theory, we derive several new crossing formulas at the two-dimensional percolation point. High-precision simulation confirms these results.  Integrating them gives  a unified derivation of Cardy's formula for the horizontal crossing probability $\Pi_h(r)$, Watts' formula for the horizontal-vertical crossing probability $\Pi_{hv}(r)$, and Cardy's formula for the expected number of clusters crossing horizontally $\mathcal{N}_h(r)$.  The main step in our approach implies the identification of the derivative of one primary operator with another. We present  operator identities that support this idea and suggest the presence of additional symmetry in $c=0$ conformal field theories.
\end{abstract}
\maketitle

%SECTION
\section{Introduction\label{intro}}
Percolation in two-dimensional systems remains under very active current study,  despite a long history.  The 2-D percolation point has been explored with a wide variety of methods, including  conformal field theory (CFT)  \cite{JC1, KSZ}, modular forms \cite{KZ}, computer simulation \cite{KSZ}, other field-theoretic methods \cite{BD}, Stochastic L\"owner Evolution (SLE) processes \cite{D} and other rigorous methods \cite{Aiz}.  (We cite only a very few representative works since the literature is so extensive.)

Crossing probabilities are of great interest in studies of the percolation point in two dimensions.  In geometries with edges, these conformally invariant quantities give the probability that percolation configurations cross between some specified set of intervals on the boundary of the system. Perhaps the best known example is Cardy's equation for the horizontal crossing probability $\Pi_h(r)$ \cite{JC1} (which was later proven rigorously for a particular realization of percolation \cite{SS}). This, the probability that a percolation cluster connects the two vertical sides of a rectangle of aspect ratio (width/length) $r$, is given by
\be\label{Pih}
\Pi_h(\l) = C\; \l^{1/3}{}_2F_1(1/3,2/3;4/3;\l) \;,
\ee
with $C = 2 \pi \sqrt{3}/\Gamma(\frac1 3)^3  
=  0.56604668\ldots$.
The cross-ratio $\l$ is related to $r$ by conformally mapping three consecutive corners of the rectangle to $1$, $\infty$ and $0$ so that the fourth corner lies on the point $\l$, with $0 \le \l \le 1$.  The interior of the rectangle maps to the upper half-plane.  Cardy used arguments of conformal field theory; primarily that the (boundary) operator which changes free to fixed boundary conditions on an edge of the system is $\psi_1  := \phi_{1,2}$ in the $c=0$ Kac table (the notation  $\psi_n := \phi_{1,n+1}$ simplifies the expressions for the boundary operator product expansion coefficients considered below \cite{JSPK}). 

The probability  $\Pi_{hv}(r)$ that all four sides of the rectangle are connected by a single percolating cluster was determined by Watts \cite{W}, using an extension of Cardy's arguments (see also the recent rigorous proof of Dub\'{e}dat \cite{D}).  This may be written as
\be \label{Pihv}
\Pi_{h v}(\l)=\Pi_h(\l)-\Pi_{h \overline{v}}(\l) \;,
\ee
where $\Pi_{h \overline{v}}$ denotes the probability of a horizontal crossing without a vertical crossing,
\be \label{Pihbarv}
\Pi_{h \overline{v}}(\l)=\frac{\sqrt 3}{2 \pi} \; \l \; {}_3F_2(1,1,4/3;5/3,2;\l) \;.
\ee
To derive this result, Watts made use of a higher-order null vector in the $c=0=h$ Verma module.
 
Finally, the expected number of clusters crossing horizontally, $\mathcal{N}_h(r)$, has also been determined by Cardy \cite{JC2,JC3} (and later via rigorous methods \cite{SS2}). This calculation involves identifying percolation as the $q \to 1$ limit of the $q$-state Potts model, and taking a derivative with respect to $q$ at $q=1$.  Maier \cite{M} pointed out that the result may be expressed as 
\bea \label{Nh1}
\mathcal{N}_h(\l)&=&\Pi_h(\l)-\frac12\Pi_{h \overline{v}}(\l) + \frac{\sqrt{3}}{4 \pi} \ln\bigg(\frac{1}{1-\l}\bigg) \;.
\eea

The motivation for this paper is the remark by Maier \cite{M} that the fifth-order differential equation  which arises from the null vector used by Watts \cite{W} to determine $\Pi_{h v}$ has, among its additional solutions \cite{KZ}, both $\Pi_{h}$ and $\mathcal{N}_h$.  This mathematical observation has, to our knowledge, eluded explanation. In this paper, using a simple adaptation of Cardy's method, we give a unified derivation of all three formulas.  In section \ref{Ceqrev} we calculate $\Pi_h$, based on a physical interpretation of the $\psi_3 := \phi_{1,4}$ operator. Section \ref{newfs} extends this method to three new crossing formulas.  Numerical verification of these results is given in section \ref{numerical}; by integrating them, section \ref{unfderv} reproduces the three known crossing quantities.  Our derivation makes use of primary operators only, avoiding higher-order null vectors and does not require reference to the Potts models to obtain $\mathcal{N}_h$.   Then, in section \ref{OpIDs}, we point out that our method implies  proportionality of $\psi_3$ and the derivative of $\psi_1$, and explore some further consequences of this identification. 

%SECTION
\section{Cardy's equation revisited\label{Ceqrev}}

In this section, we briefly review Cardy's derivation of the horizontal crossing probability $\Pi_h(\l)$, and then present the approach used to derive it here, as an introduction to the more interesting results below.

In \cite{JC1} Cardy determined $\Pi_h(\l)$ via the four-point function $ \langle \psi_1(0) \psi_1(\l) \psi_1(1) \psi_1(\infty) \rangle$.  Here, adjacent pairs of operators mark the intervals $(0,\l)$ and $(1,\infty)$ between which  the crossing occurs.

\begin{figure} 
\centering
%\begin{center}
%\includegraphics*[bb = 0 0 600pt 300pt,scale=.5]{config2.eps}
\includegraphics[width = 3.0in]{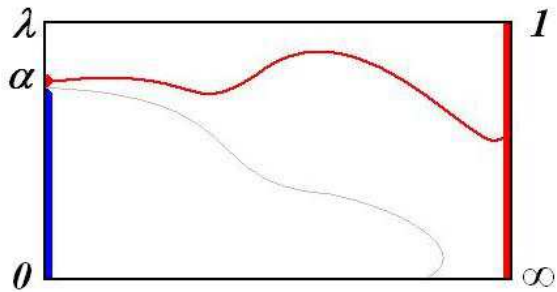}
%\end{center}
\caption{Effects of $\psi_3(\a)$--configurations contributing to $\Pi_{h;\a}$. Crossing path shown in red, dual path in grey.  Thick (thin) boundary lines represent fixed (free) edges.} \label{config}
\end{figure}

Our figures herein are shown as rectangles, while our formulas are given in terms of upper half-plane variables e.g.\ $\l$. These two geometries are equivalent under a conformal mapping. One visualizes crossings in rectangles for consistency with  common usage (e.g.\ ``horizontal crossing"), but takes  parameters to lie on the real line  for mathematical simplicity. Thus  ``the $(0,\l)$ edge of the rectangle" in fact indicates the interval on the real axis that maps into the corresponding side of the rectangle.  Figure \ref{config} illustrates how the four points $(0, \l , 1, \infty)$ on the real axis map to the rectangle.

Cardy's derivation (see  \cite{JC1} or \cite{JC3} for more details) focuses on the comparison of the two possible fixed boundary condition assignments; either the same or different. For percolation, a fixed boundary either allows clusters to touch it or not; thus a rectangle with two fixed vertical edges and free horizontal edges either includes all clusters, or excludes horizontally crossing clusters. Therefore, by inserting a $\psi_1$ (which changes the boundary condition from fixed to free) at each of the four corners of the rectangle, and considering the second-order differential equation implied by their null vector, one finds two solutions, which may be taken to be $1$ and $\Pi_h(\l)$.  
Thus
\be \label{Pihcfs}
\Pi_h(\l) = \langle \psi_1^{a f}(0)  \psi_1^{f a}(\l)  \psi_1^{a f}(1) \psi_1^{f a}(\infty) \rangle - \langle \psi_1^{a f}(0)  \psi_1^{f b}(\l)  \psi_1^{b f}(1) \psi_1^{f a}(\infty) \rangle \;.
\ee
Here, the superscripts indicate the boundary condition change; $f$ denoting free and $a$ or $b$ fixed boundary conditions, with $a \ne b$.  Thus the first term includes all configurations, and is a constant, independent of $\l$, while the second removes those configurations with no horizontal crossing.   If we normalize our (boundary) operators so that $\langle \psi_i(0) \psi_i(x) \rangle = x^{-2h_i}$, it follows that the two sides of (\ref{Pihcfs}) are equal, with no multiplicative constant, and (\ref{Pihcfs}) becomes
\be
\Pi_h(\l)=1-\Pi_h(1-\l) \; ,
\ee
which may also be derived using duality (see \cite{JC1,JC3} for more details on these matters).

Our method modifies the standard derivation as follows. 
Consider the probability density $\Pi_{h;\a}$ that the crossing connects  the interval $(\a, \a+\mathrm{d}\a)$ but {\bf not} the interval $(0,\a)$ to the interval $(1,\infty)$, where $0 \le \a \le \l \le 1$. This is
\be \label{dPih}
\Pi_{h;\a}\: \mathrm{d}\a=\Pi_h(\a+\mathrm{d}\a)-\Pi_h(\a)\quad \Rightarrow \quad \Pi_{h;\a}= \partial_{\a}\Pi_h(\a)
\ee
The configurations that will contribute to this probability are those which have a percolation cluster connecting the point $\a$ to the interval $(1,\infty)$ and also have, on the fixed boundary side of that percolation cluster, a dual path from $\a$ to $(-\infty,0)$, as illustrated in Figure \ref{config}.  Here, the fact that the small interval is not connected to $(0,\a)$ ensures the presence of the dual path; and the differentiation in (\ref{dPih}) removes the constant term in (\ref{Pihcfs}), so that a crossing cluster must attach to $\a$.

At first sight, it might seem that  $\Pi_{h;\a}$ should also depend on $\l$.  However, this quantity, which is specified in the half-plane, can be mapped to a rectangle with any aspect ratio $r$, as mentioned.  When this is done, the length of the image of the interval $(0,\a)$  will vary according to $\l$, which also determines $r$.

Now the operator expected  \cite{Aiz,JSPK} to generate a percolation cluster and dual path should have dimension $h=1$.  This suggests that it is $\psi_3 :=\phi_{1,4}$.   For the moment we simply assume this, and explore its consequences.  In section \ref{OpIDs} we  give a better justification (and consider its implications). 

Note that $\psi_3$, since it arises in the operator product expansion of three $\psi_1$ operators, must sit at a fixed-free boundary change, as shown in Figure \ref{config}.  

Therefore we have
\be \label{Piha}
\Pi_{h;\a} = K\; \langle \psi_1(0) \psi_3(\a)  \psi_1(1) \psi_1(\infty) \rangle \;,
\ee
where $K$ is a constant.  If we set
\be\label{Kval1}
K=\frac{{3^{1/4}}}{2\; \sqrt{\pi}} \;,
\ee
it turns out that (\ref{Piha}) will be properly normalized, as shown below.
In section \ref{OpIDs} we justify (\ref{Kval1}) directly, without reference to percolation, by means of the operator product expansion.  Note that $K^2$ is exactly $1/2$ the constant appearing in (\ref{Pihbarv}).

Thus
\bea \label{Pihnew}
\nonumber\Pi_h(\l)&=&\int_0^{\l}\Pi_{h;\a}\;\textrm{d}\a\\
\label{intPih}&=&K \int_0^{\l}\langle \psi_1^{f a}(0) \psi_3^{a f}(\a)  \psi_1^{f b}(1) \psi_1^{b f}(\infty)  \rangle\;\textrm{d}\a \;.
\eea
Therefore we must determine the correlation function  (\ref{Piha}). Now we may write
\be \label{pihcf}
\langle \psi_1(x_4)\psi_1(x_3)\psi_3(x_2)\psi_1(x_1) \rangle =
\frac{(x_4-x_1)(x_4-x_3)}{(x_3-x_1)(x_4-x_2)^2}F\left( \frac{(x_2-x_1)(x_4-x_3)}{(x_3-x_1)(x_4-x_2)} \right) \;.
\ee
Since $\psi_1$ is, as mentioned, a level-two operator, the space of possible solutions for (\ref{pihcf}) is two-dimensional.  However, in the operator product expansions of $\psi_1 \psi_1$ and $\psi_1 \psi_3$, the only common term is $\psi_2 := \phi_{1,3}$, so only one conformal block enters; i.e. the solution space is one-dimensional.  To determine it, we apply the null state condition to two different  $\psi_1$ operators in (\ref{pihcf}).  This gives two different second-order differential equations for $F$. Subtracting them so as to cancel the highest-order term gives 
\be
0=F'(\a) + \frac{2(1-2\a)}{3\a(1-\a)}F(\a) \;.
\ee
This equation fixes the single conformal block as
\be \label{1blk}
\mathcal{F}_{1 3, 1 1}^2 (\a) = \left( \a(1-\a) \right)^{-2/3}\; ,
\ee
where the superscript $2$ refers to $\psi_2 := \phi_{1,3}$ which appears in the operator product expansion of $\psi_1$, both with itself and with  $\psi_3$. (Our conformal blocks are normalized so that  $\mathcal{F}_{i j, k l}^n (x) \sim x^{h_n-h_i-h_j}$.) This leads to 
\be\label{CF1311}
\langle \psi_1^{f a}(0) \psi_3^{a f}(\a)  \psi_1^{f b}(1) \psi_1^{b f}(\infty)  \rangle = C_{1 2 3}C_{1 1 2} \left( \a(1-\a) \right)^{-2/3} \;,
\ee
where the usual superscripts (indicating the boundary conditions) on the  boundary operator product expansion coefficients $C_{i j k}$ \cite{L} have been suppressed, as a consequence of duality  \cite{JSPK}.  
Inserting this correlation function into (\ref{Pihnew}) then reproduces (\ref{Pih})
\bea
\nonumber\Pi_h(\l)&=&K  C_{1 2 3}C_{1 1 2} \int_0^{\l} \a^{-2/3}(1-\a)^{-2/3} \textrm{d}\a\\ \nonumber
&=&3 K  C_{1 2 3}C_{1 1 2} \; \l^{1/3}\;{}_2F_1(1/3,2/3;4/3;\l)\\
&=&C \; \l^{1/3}\;{}_2F_1(1/3,2/3;4/3;\l) \;, \label{PihC}
\eea
where we have made use of  (\ref{Kval1}) as well as  $C_{1 2 3}=2 \sqrt2 \pi/3\;\Gamma[1/3]^{3/2}$ and $C_{1 1 2}=\sqrt{2 \pi}\; 3^{1/4}/\Gamma[1/3]^{3/2}$ \cite{JSPK}.  

The function $\pi_{h;\a} := \p_{\a} \Pi_h(\a)$ is a simple example of what we call a \emph{first crossing density}.    The term ``first'' indicates a probability density for configurations that, when we start at the origin and move towards the point $\l$, first contain a crossing cluster in the neighborhood of $\a$.  Herein, the lower case ($\pi$) distinguishes crossing probability \emph{densities} from crossing \emph{probabilities}, represented with upper case ($\Pi$).

With only one $\psi_3$ in the correlation function  we reproduce Cardy's result for $\Pi_h(\l)$.  However by inserting an additional $\psi_3$ operator we can generate more complicated first (and other) crossing densities.  These then give a new derivation of Watts' equation for $\Pi_{h v}$  \cite{W}, and Cardy's expression for the mean number of horizontal crossing clusters $\mathcal{N}_h(\l)$  \cite{JC2,JC3}, as well as $\Pi_h$.

%SECTION
\section{New crossing formulas\label{newfs}}

\begin{figure}[ht]
\centering
\includegraphics[width = 6.0in]{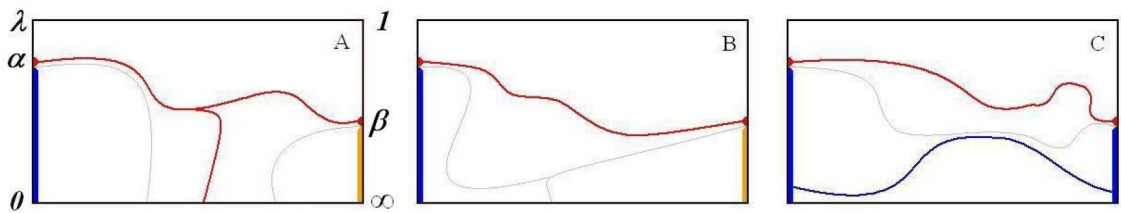}
\caption{Configurations consistent with $\langle \psi_1(\infty) \psi_3(\b) \psi_3(\a) \psi_1(0) \rangle$. \label{config3}}
\end{figure}

In order to find new results, we consider the correlation function $\langle \psi_1(\infty) \psi_3(\b) \psi_3(\a) \psi_1(0) \rangle$ with  $0<\a<\l$, and $1<\b$.  By a simple extension of the argument above, one sees that there are three configurations  consistent with this function, illustrated  in figure \ref{config3}.

Let $\pi_h^b(\a, \b)$ ($\pi_h^{\bar b}(\a, \b)$) denote the first crossing probability density for configurations of type $A$ ($B$); first crossings from $\a$ to $\b$ that also make (do not make) contact with the bottom edge of the rectangle.  

Similarly, $\nu_h(\a,\b)$ denotes the crossing density of configurations of type $C$.  Now $\nu_h(\a,\b)$  is not a {\it first} crossing density;  rather it includes configurations with crossings from $\a$ to $\b$ that are not the first crossing, but are distinct from previous crossings--hence the notation $\nu$ in place of $\pi$.  Thus configurations with multiple crossings contribute to $\nu_h(\a,\b)$ for each pair of values $\a$ and $\b$ spanned by a new cluster.  Integrating it therefore  counts configurations with $n$ horizontal crossings $n-1$ times. We use this below to calculate $\mathcal{N}_h(\l)$.

Now the correlation function 
\be
\langle \psi_1(x_4) \psi_3(x_3) \psi_3(x_2) \psi_1(x_1) \rangle=
\left( \frac{x_4-x_1}{(x_4-x_2)(x_3-x_1)}\right)^2 F\left( \frac{(x_2-x_1)(x_4-x_3)}{(x_3-x_1)(x_4-x_2)} \right)\; .
\ee
It follows that
\be
\langle \psi_1(\infty) \psi_3(\b) \psi_3(\a) \psi_1(0) \rangle = \b^{-2}F(\a/\b)\; .
\ee

Utilizing the second-order null vector for $\psi_1$  we find 
\be
0=F''(x)+\frac{2(1-8 x)}{3x(1-x)}F'(x)-\frac{2(1-6 x^2)}{3 x^2(1-x)^2}F(x) \;.
\ee
Solving  and selecting the appropriate conformal blocks gives
\bea
\mathcal{F}_{1 3, 3 1}^2(x) &=& \frac{1+x}{(1-x)^{5/3}x^{2/3}}\\
\mathcal{F}_{1 3, 3 1}^4(x)&=&\frac{5(1+2 x-(1-x^2)\,{}_2F_1(1,4/3,5/3,x))}{6(1-x)^2} \label{blk4}\\
\mathcal{F}_{3 3, 1 1}^0(1-x)&=&\frac{(1+2 x+(1-x^2)\,{}_2F_1(1,4/3,5/3,1-x))}{3(1-x)^2}\\
\mathcal{F}_{3 3, 1 1}^2(1-x)&=&\frac{1+x}{2(1-x)^{5/3}x^{2/3}} \;,
\eea
with superscripts defined as in (\ref{1blk}). The crossing symmetry relations for these conformal blocks follow using hypergeometric identities \cite{AbSt} for $x \to 1-x$, and may be written using the operator product expansion  coefficients $C_{1 2 3}$ and $C_{1 1 2}$ quoted above; we also make use of $C_{2 3 3}=2^{7/2}  \pi^{3/2}/3^{9/4}\Gamma[1/3]^{3/2}$ and $C_{1 3 4}=\sqrt{2/5}$  \cite{JSPK} (note that $K= C_{1 1 2}/3 C_{1 2 3}$, see (\ref{Kval1})).  (We have explicitly verified that the hypergeometric identities are consistent with these values.) Thus
\bea
\label{denXsym1}C_{1 2 3}^2 \mathcal{F}_{1 3, 3 1}^2(x) &=& C_{1 1 2}C_{2 3 3}\mathcal{F}_{3 3, 1 1}^2(1-x)\\
\label{denXsym2}C_{1 3 4}^2\mathcal{F}_{1 3, 3 1}^4(x) &=& \mathcal{F}_{3 3, 1 1}^0(1-x)-C_{1 1 2}C_{2 3 3}\mathcal{F}_{3 3, 1 1}^2(1-x)\\
\label{denXsym3}\mathcal{F}_{3 3, 1 1}^0(1-x) &=&C_{1 2 3}^2 \mathcal{F}_{1 3, 3 1}^2(x) +C_{1 3 4}^2\mathcal{F}_{1 3, 3 1}^4(x) \;.
\eea

Given these blocks, we may use the boundary conditions to determine which configurations they correspond to.
Fixing both intervals $(0,\a)$ and $(\b, \infty)$ in the same way determines the conformal block, so that
\be \label{F0blk}
\langle \psi_1^{f a}(\infty) \psi_3^{a f}(\b) \psi_3^{f a}(\a) \psi_1^{a f}(0) \rangle = \b^{-2}\mathcal{F}_{3 3, 1 1}^0(1-\a/\b) \;.
\ee 
With the same boundary condition on these two intervals  none of the configurations in Figure \ref{config3} are excluded.  Thus (\ref{F0blk}) is proportional to the sum of all three crossing densities.  

Multiplying by  $K^2$ (see (\ref{Kval1})) again results in proper normalization, as explained below.  Hence 
\be\label{XdensResABC}
\pi_h^b(\a, \b) + \pi_h^{\bar b}(\a, \b)+\nu_h(\a, \b) = K^2 \b^{-2}\mathcal{F}_{3 3, 1 1}^0(1-\a/\b) \;.
\ee

On the other hand, fixing the two intervals $(0, \a)$ and $(\b, \infty)$ differently leads to
\be
\langle \psi_1^{f b}(\infty) \psi_3^{b f}(\b) \psi_3^{f a}(\a) \psi_1^{a f}(0) \rangle = C_{1 1 2} C_{2 3 3} \b^{-2}\mathcal{F}_{3 3, 1 1}^2(1-\a/\b) \;.
\ee 
In this case configurations of type $C$ are excluded so that
\be\label{XdensResAB}
\pi_h^b(\a, \b) + \pi_h^{\bar b}(\a, \b) = K^2 C_{1 1 2}C_{2 3 3} \beta^{-2}\mathcal{F}_{3 3, 1 1}^2(1-\a/\b)\; .
\ee

Using (\ref{denXsym2}), (\ref{XdensResABC}) and (\ref{XdensResAB})  we can now find the crossing density
\bea
\nu_h(\a, \b) &=& K^2\b^{-2}\left( \mathcal{F}_{3 3, 1 1}^0(1-\a/\b) - C_{1 1 2}C_{2 3 3}\mathcal{F}_{3 3, 1 1}^2(1-\a/\b) \right)\\
&=& K^2 C_{1 3 4}^2 \b^{-2}\mathcal{F}_{1 3, 3 1}^4(\a/\b)\; .
\eea

To separate $\pi_h^b(\a, \b)$ and $\pi_h^{\bar b}(\a, \b)$ we fix the boundary conditions on the bottom edge $(-\infty, 0)$ to differentiate first crossings that touch the bottom edge (type A) and those that do not (type B or C).

The two-point function
\be
\langle \psi_3^{a f}(\a) \psi_3^{f a}(\b) \rangle\ = (\b-\a)^{-2} \;
\ee
includes clusters connecting $\a$ and $\b$, but  not touching the bottom edge, since it is part of a single fixed interval isolated by a dual path.  This excludes crossings of type $A$, so that
\be
\pi_h^{\bar b}(\a,\b) + \nu_h(\a, \b) = K^2(\b-\a)^{-2}\; .
\ee
This leads to 
\bea
\pi_h^b(\a,\b)&=& K^2 \left( \b^{-2} \mathcal{F}_{3 3, 1 1}^0(1-\a/\b)-(\b-\a)^{-2}\right) \\
\pi_h^{\bar b}(\a,\b) &=& K^2 \left( (\b-\a)^{-2}-C_{1 3 4}^2 \b^{-2} \mathcal{F}_{1 3, 3 1}^4(\a/\b) \right) \;.
\eea

Collecting and simplifying these  results  gives the formulas
\bea\label{pihb}
\pi_h^b(\a, \b) &=& \frac{(\b+\a)\;{}_2F_1(1,4/3,5/3,1-\a/\b)-2\b}{4\pi\sqrt{3}\; \b^2(\b-\a)} \\
\label{pihbarb}
\pi_h^{\bar b}(\a, \b) &=& \frac{(\b+\a)\;{}_2F_1(1,4/3,5/3,\a/\b)+2\b}{4 \pi \sqrt{3}\; \b^2(\b-\a)} \\
\label{nuh}
\nu_h(\a, \b) &=& \frac{\b^2+2\a \b-(\b^2-\a^2)\;{}_2F_1(1,4/3,5/3,\a/\b)}{4 \pi \sqrt{3}\; \b^2(\b-\a)^2} \;.
\eea
These results are new, to our knowledge.  They are sufficient to reproduce all three previously known crossing quantities, as we now proceed to demonstrate.  It is interesting that only a single  (${}_2F_1$) hypergeometric function enters.

%SECTION
\section{Numerical verification\label{numerical}}

To verify these results, we carried out simulations using hull walks on a square system, for bond percolation on the square lattice,  where $p_c = 1/2$.  For this system, a hull walk is a simple walk at $45^\circ$ to the bonds that turns left or right with equal probability at each step, except when it encounters a site previously visited, in which case it always turns to avoid retracing its path.  In this way the walk lays down the adjacent occupied and vacant bonds of a hull for the percolating system \cite{ZCS,G}.  We tested the functions $\pi_h^b(\alpha,\beta)$, $\pi_h^{\bar b}(\alpha,\beta)$, and $\nu_h(\alpha,\beta)$ for the half-plane transformed to a square system of side length $1$, with $\alpha$ chosen to correspond to  $(x,y) = (0, 1/2)$ (the mid-point on the left-hand boundary), and $1 \le \b \le \infty$, so the corresponding point varies along the right-hand boundary.  

\begin{figure}
\centering
\includegraphics[width = 4.0in]{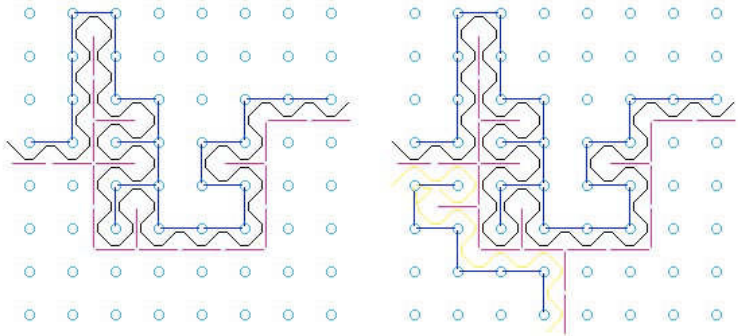}
\caption{Hull-generating walk algorithm to check for crossing densities on an $8 \times 8$ lattice.  (left) A hull (the black curve) corresponding to a crossing cluster that does not touch left or bottom boundaries, and does not cross the top boundary.  (right) Second hull walk (shown in yellow, below the first) showing that there are no lower crossing clusters.  Blue segments: occupied bonds on the lattice.  Red segments: bonds on the dual lattice, corresponding to vacant bonds on the original lattice.   \label{walkfig}}
\end{figure}

In the simulation, the walk was started on the left-hand side of the square at the point 
$(0,1/2)$.   The requirement that the hull borders a first-crossing cluster starting at that point means that the
hull cannot touch anywhere on the entire left-hand side.  Similarly, if the hull crosses the top boundary, then the trial is terminated, since that event corresponds to the vacant bonds of the hull touching the top, preventing a horizontal crossing.   Walks that touch the lower boundary (indicating the
cluster of occupied bonds touches that boundary) were allowed to continue.  Those walks that touch the bottom and continue to the cross to the right-hand side contribute to $\pi_h^b$; those crossing ones that don't touch the bottom were further checked for crossing clusters below them.  To do this,  a second hull walk was started from $(0, 1/2)$, to represent the hull of the dual crossing, as in the shaded dual-lattice paths shown in Figure \ref{PihvbFig}.  If this walk intersects the bottom,  there cannot be any horizontal crossing clusters below the first simulated crossing cluster, and the walk contributes to $\pi_h^{\bar b}$. Otherwise, if it doesn't touch the bottom before it crosses, there must be at least one lower horizontal crossing, and the event contributes to $\nu_h$.

Figure \ref{walkfig} shows an example of a walk on a system of $8 \times 8$ bonds.  Here the blue circles are lattice vertices, and the blue edges  occupied bonds on the lattice.  The red edges are occupied bonds on the dual lattice, corresponding to vacant bonds on the original lattice.  The figure on the left shows, in black, the walk corresponding to a crossing cluster.  The system is prepared by setting one vacant bond (or a dual-lattice bond, red) immediately below the lattice point corresponding to $(0, 1/2)$, so that the walk is guaranteed to enter the system and the point $y = 1/2$ will be at the boundary between occupied and vacant bonds. (Walks that exit the system at the entry point are discarded.)  The walk then generates the remaining occupied and vacant bonds of the hull.  This particular walk terminated when it intersected the right boundary.  Next, to check if it was a {\it first-crossing} cluster, a second hull was initiated, starting on the vacant (or dual-lattice) bond in the first column.  This hull is shown in yellow,  in the figure on the right.  To keep the second walk from leaving the system on the left, we added occupied bonds in the lower first column.  This particular walk reached the bottom before reaching the right-hand side, indicating that there were no additional crossing clusters below the first crossing cluster.

Because this method generates only the hull of the cluster, and simultaneously yields the type of crossing, it is very efficient.  In several days of computer time, we were able to generate  $3.3  \times  10^{11}$ hulls on a lattice of $512 \times 512$ bonds.   Only $0.0016505$ of the walks succeeded in making it across without hitting the top or left-hand sides.  Of these, a fraction $0.6456$ hit the bottom and contributed to $\pi_h^b$, while the remaining $0.3544$ crossed without hitting the bottom.  Of the latter, a fraction $0.92982$ did not have additional clusters below them (contributing to $\pi_h^{\bar b}$) and
$0.07018$ did (contributing to $\nu_h$).  In all, only a fraction $0.0001707$ of all initiated walks
corresponded to events that contribute to $\nu_h$.

The above fraction of multiple crossing events, $0.07018$, is somewhat above the predicted value $0.069189$, which is found by integrating the formulas for $\nu_h$ and $\pi_h^{\bar b}$ and taking the ratio of the integral of the former to the sum of the integrals of the former and latter.  This difference can be attributed to finite-size effects, which is apparent by considering this quantity for lattices of side length $L=64\; (0.07653)$, $128\; (0.07294)$, $256\; (0.07108)$, and $1024\; (0.06978)$.  The data fit very well to a straight line when plotted as a function of $1/L$, with an intercept of $0.06928$, quite close to the predicted value.

\begin{figure}
\centering
\includegraphics[width = 6.0in]{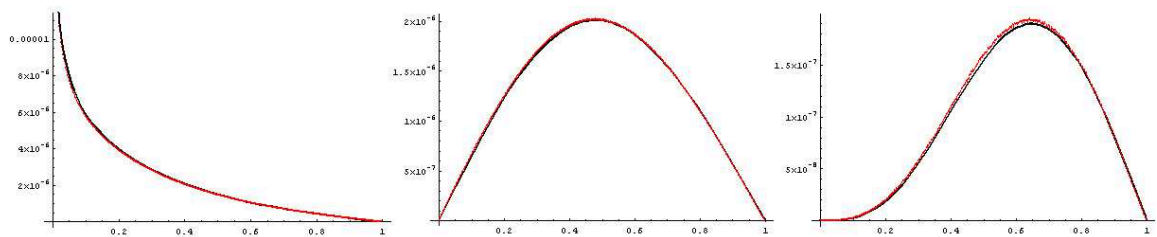}
\caption{Comparison of simulation (red dots) on a $512 \times 512$ lattice with theory (Equations (\ref{pihb}), (\ref{pihbarb}), and (\ref{nuh})).\label{JakesFig}}
\end{figure}

In Figure \ref{JakesFig} we compare the numerical results with the theory. The data are  plotted vs.\ the position of the point on the right-hand side corresponding to $\b$, where on the left-hand side we pick the mid-point, as mentioned above. The continuum coordinate was taken to be $y=(Y+1/2)/512$, where the lattice coordinate $Y=0,1, \ldots 511$.    The relative difference between the two curves is on the order of $2 \%$, except near the corners of the square and where the functions are small, in which case the difference is somewhat larger.  The overall deviation in $\nu_h$ compared with the theory is also a finite-size effect which extrapolates  nearly to zero when $L \to \infty$.
There is also a slight bias to our results reflecting the fact that a finite system is not perfectly symmetric with respect to the boundary conditions of the walk entering and leaving the system.  We have found that this bias also diminishes as the system size increases. 

In conclusion, we find very good agreement between simulations and theory for these various quantities.

%SECTION
\section{Unified derivation of crossing formulas\label{unfderv}}

Next  we integrate our formulas, to  re-derive the known results for the horizontal crossing probability $\Pi_h$, the horizontal-vertical crossing probability $\Pi_{h v}$, and the expected number of horizontal crossing clusters $\mathcal{N}_h(\l)$.

Now $\Pi_h^b(\l)$, the probability that there exists a horizontal crossing cluster that also touches the bottom edge of the rectangle (such as the one illustrated in Figure \ref{config3}[A]), is given by 
\bea
\nonumber \Pi_h^b(\l) &=& \int_0^{\l} \int_1^{\infty} \pi_h^b(\a, \b) \mathrm{d}\b\; \mathrm{d}\a\\
&=&\int_0^{\l} \int_1^{\infty} \frac{(\b+\a)\;{}_2F_1(1,4/3,5/3,1-\a/\b)-2\b}{4\pi\sqrt{3}\; \b^2(\b-\a)}\; \mathrm{d}\b\; \mathrm{d}\a \;.
\eea
(Note that there can only be one such cluster in any configuration, so $\Pi_h^b(\l)$ is also the expected number of this type of cluster.)
Next let $\b \to \a/\xi$, so that
\be
\Pi_h^b(\l) = \int_0^{\l} \frac{1}{4 \pi \sqrt{3}\; \a} \int_0^{\a} \frac{(1+\xi)\;{}_2F_1(1,4/3,5/3,1-\xi)-2}{(1-\xi)}\; \mathrm{d}\xi\; \mathrm{d}\a \;,
\ee
then transform the hypergeometric function with the same identities used in deriving the crossing symmetries (\ref{denXsym1}-\ref{denXsym3}).  This gives
\bea \nonumber
\Pi_h^b(\l) &=& \int_0^{\l} \frac{C_{1 1 2}^2}{9 \a} \int_0^{\a} \frac{1+\xi}{(1-\xi)^{5/3}\xi^{2/3}}\; \mathrm{d}\xi\; \mathrm{d}\a\\
&& - \int_0^{\l} \frac{1}{4 \pi \sqrt{3}\; \a} \int_0^{\a} \frac{(1+\xi)\;{}_2F_1(1,4/3,5/3,\xi)+2}{(1-\xi)}\; \mathrm{d}\xi\; \mathrm{d}\a \label{nicestep} \;.
\eea
(The coefficient of the first integral is given in terms of $C_{1 1 2}^2$ for reasons that will be clear shortly.)  The identity
\be\label{randHypGeoIdent}
\p_\xi \left( 3 \xi \;{}_2F_1(1,4/3,5/3,\xi)\right) = \frac{(1+\xi)\;{}_2F_1(1,4/3,5/3,\xi)+2}{(1-\xi)}
\ee
follows from the integral representation of the hypergeometric function. Using it in (\ref{nicestep}) leads to
\bea \nonumber
\Pi_h^b(\l) &=& \int_0^{\l} \frac{C_{1 1 2}^2}{3 \a^{2/3}(1-\a)^{2/3}}\; \mathrm{d}\a\\
&&-\frac{\sqrt{3}}{4 \pi} \int_0^{\l}\;{}_2F_1(1,4/3,5/3,\a)\; \mathrm{d}\a \;.
\eea
By (\ref{PihC}),  the first term equals $\Pi_h(\l)$.  To evaluate the second integral we use the identity
\be
\p_{\a}\left( \a\;{}_3F_2(1, 1, 4/3; 5/3, 2; \a) \right)=\;{}_2F_1(1,4/3,5/3,\a) \;,
\ee
which is easily derived from the series for the hypergeometric function.  The final result is
\bea
\nonumber \Pi_h^b(\l) &=& \Pi_h(\l)-\frac{\sqrt{3}}{4 \pi} \l\;{}_3F_2(1, 1, 4/3; 5/3, 2; \l)\\
\label{Pihb}&=& \Pi_h(\l)-\frac{1}{2}\Pi_{h \bar v}(\l) \; ,
\eea
where we have made use of (\ref{Pihbarv}). 

The treatment for $\Pi_h^{\bar b}(\l)$, the probability of horizontal crossing when the lowest spanning cluster does {\it not} touch the bottom, follows analogously.  Integrating $ \pi_h^{\bar b}(\a, \b)$ over $\a$ and $\b$ as above, we arrive at the second term in (\ref{nicestep}). Thus
\be\label{Pihbarb}
\Pi_h^{\bar b}(\l) = \frac{1}{2} \Pi_{h \bar v}(\l) \;.
\ee

Equations (\ref{Pihb}) and (\ref{Pihbarb}) allow us to derive $\Pi_{h v}(\l)$.  The configurations  that contribute to $\Pi_h^{\bar b}(\l)$ (Figure \ref{config3}[B]) are such that the crossing from $\a$ to $\b$ is the first \emph{and} does not touch the bottom edge. Thus the dual path from $\a$ to $\b$ must itself touch the bottom edge.  Therefore, by duality, $\pi_h^{\bar b}(\a, \b)$ is the probability density of a horizontal crossing that touches the bottom but is separated from the top by a dual cluster from $\a$ to $\b$.  Thus $\Pi_h^{\bar b}(\l) = \Pi_{h \bar v}^b(\l)$, where $\Pi_{h \bar v}^b(\l)$ is the probability of a horizontal crossing cluster that touches the bottom, but is prevented from crossing vertically by a horizontal dual path.

Finally, $\Pi_{h v}(\l)$ is the probability of a horizontal crossing cluster that touches both the top and bottom. Hence
\bea
\nonumber \Pi_{h v}(\l) &=&\Pi_h^b(\l)-\Pi_{h \bar v}^b(\l) = \Pi_h^b(\l)-\Pi_h^{\bar b}(\l)\\
\label{Pihvexp}&=& \Pi_h(\l) - \Pi_{h \bar v}(\l) \;.
\eea
Thus, by integrating and combining our new first crossing densities, we arrive at Watts' equation (\ref{Pihv}) for the horizontal-vertical crossing probability.

\begin{figure}
\centering
\includegraphics[width = 5.0in]{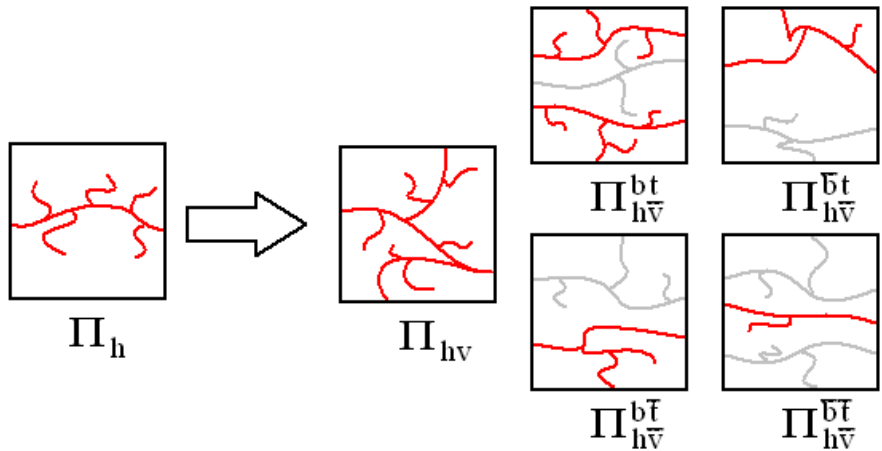}
\caption{The five distinct configurations that contribute to $\Pi_h$.  Paths in clusters are red, dual paths grey. 
\label{PihvbFig}}
\end{figure}

Equations  (\ref{Pihb}) and  (\ref{Pihbarb})  can also be derived by a duality argument, which is a non-trivial check of our results.  To do this, extend our notation, as shown in Figure \ref{PihvbFig}.
The $b$ and $t$ ($\bar b$ and $\bar t$) superscripts denote configurations for which there is a horizontal crossing cluster which touches (does not touch) the bottom or top edge of the rectangle respectively.    The four rightmost diagrams in Figure \ref{PihvbFig} include all the configuration types consistent with $\Pi_{h \bar v}$.

Thus
\bea \nonumber
\Pi_{h \bar v}&=&\Pi_{h \bar v}^{b \bar t}+\Pi_{h \bar v}^{\bar b t}+\Pi_{h \bar v}^{b t}+\Pi_{h \bar v}^{\bar b \bar t},\; \\ 
\nonumber \Pi_h^{ \bar b}&=&\Pi_{h \bar v}^{\bar b t}+\Pi_{h \bar v}^{\bar b \bar t}, \; \mathrm{and} \\
 \nonumber \Pi_h^{b}&=&\Pi_{hv}+\Pi_{h \bar v}^{b \bar t}+\Pi_{h \bar v}^{b t} \;. 
\eea
But by duality $\Pi_{h \bar v}^{b \bar t}=\Pi_{h \bar v}^{\bar b t}$ and $\Pi_{h \bar v}^{b t}=\Pi_{h \bar v}^{\bar b \bar t}\;$, from which (\ref{Pihb}) and  (\ref{Pihbarb}) follow.

Finally, we derive the expected number of horizontal crossing clusters using $\nu_h(\a, \b)$.  Recall that this density gives the probability that there is a new cluster spanning from $\a$ to $\b$ that is \emph{not} the lowest crossing cluster in the rectangle.  Thus integrating it gives a contribution of $n-1$ for each configuration with $n$ crossing clusters.  Therefore 
\bea
\nonumber \mathcal{N}_h(\l) - \Pi_h(\l) &=&\int_0^{\l} \int_1^{\infty} \nu_h(\a, \b) \mathrm{d}\b\; \mathrm{d}\a\\
\nonumber &=&\int_0^{\l} \int_1^{\infty} \left(\frac{\sqrt{3}}{4 \pi\;(\b-\a)^2}­-\pi_h^{\bar b}(\a, \b)\right) \mathrm{d}\b\; \mathrm{d}\a\\
&=&\frac{\sqrt{3}}{4 \pi} \log \left( \frac{1}{1-\l} \right)-\frac{1}{2}\Pi_{h \bar v}(\l) \;,
\eea
giving (\ref{Nh1}).

This concludes our derivation of the crossing formulas.  As mentioned, by exploiting our new crossing results, we obtain all three known results without reference to the $q$-state Potts model or use of higher-order null vectors. Next, we consider  our use of $\psi_3$ above from an operator point of view, and examine some of its consequences.

%SECTION
\section{Operator identities\label{OpIDs}}

In this section, we first consider our use of the $\psi_3$ operator in sections \ref{Ceqrev} and \ref{newfs}, and then present a calculation of the constant $K$ used to normalize our densities (see (\ref{Piha}), (\ref{Kval1}), and section \ref{newfs}). 

To begin, consider (\ref{Piha}), which, in light of (\ref{Pihcfs}) and  (\ref{dPih}), can be interpreted as replacing $\partial_z \psi_1(z) $ by $K \psi_3(z)$. Now generally, this would not be possible, since the derivative of a primary operator is not primary itself.  However the derivative of a primary operator of weight zero (like $\psi_1$) is indeed primary.

Next, (\ref{Pihcfs})  gives
\bea
\nonumber\p_{\a}\Pi_h(\a) &=& \p_{\a} \langle  \psi_1(0) \psi_1(\a) \psi_1(1) \psi_1(\infty)\rangle\\
&=&\langle  \psi_1(0) L_{-1}\psi_1(\a) \psi_1(1) \psi_1(\infty)\rangle \;.
\eea
Now the weight of $L_{-1}\psi_1$ is  $1$, the same as for $\psi_3$.  More importantly, the null operator for $\psi_3 := \phi_{1,4}$ is
\be
\mathcal{D}_{1,4} = 3L_{-1}^4-20 L_{-2}L_{-1}^2+24L_{-2}^2+24L_{-3}L_{-1}-24L_{-4}\; .
\ee
(Here  $\mathcal{D}_{r,s}$ denotes the null operator for the $\phi_{r,s}$ Kac operator.) Further,  by the $L_m$ commutation relations  (for $c=0$) one has
\bea
\nonumber \mathcal{D}_{1,4}L_{-1}  &=& (L_{-1}^3-6L_{-2}L_{-1}+6L_{-3})(3L_{-1}^2-2L_{-2}), \; \; \mathrm{i.e.}\\
\mathcal{D}_{1,4} \mathcal{D}_{1,1} &=& \mathcal{D}_{3,1} \mathcal{D}_{1,2} \;. 
\label{wattsvec}
\eea
The right hand side is exactly the level five null operator used by Watts \cite{W}!  Since $\mathcal{D}_{1,2} = 3L_{-1}^2-2L_{-2}$ is the null operator for $\psi_1$, so is $\mathcal{D}_{1,4}\mathcal{D}_{1,1}$  as well.

Therefore the weight of $L_{-1}\psi_1$ equals that of $\psi_3$, and they both obey the same null state.  Thus correlation functions involving them obey the same differential equations, and the solutions must overlap.  Hence we posit
\be \label{opid}
L_{-1}\psi_1(x) = K \psi_3(x)\; .
\ee
In section \ref{furdisc}, we discuss implications of this equation.  For the moment, consider the question as to where in the above it actually makes a difference, i.e., if we were to differentiate a correlation function containing $ \psi_1$ instead of substituting $K\psi_3$ for it, what would change?  It is easy to see that the results of section \ref{Ceqrev} would be the same; however a crucial difference occurs for  (\ref{blk4}).  Here the conformal block $\mathcal{F}^{4}$, which contributes to  $\pi_h^b$, $\pi_h^{\bar b}$ and $\nu_h$, would not appear, and our calculations would not be valid.

Now we determine the constant $K$ by comparing leading terms in the operator product expansions
\bea
\nonumber (L_{-1} \psi_1(x)) \psi_1(0) &=& \p_x \psi_1(x) \psi_1(0)\\
\nonumber &=& \p_x( \mathbf{1}(0)+\frac15 x^2 T(0) +\dots+ C_{1 1 2} x^{1/3} \psi_2(0) +\dots)\\
&=&\frac25 x T(0)+\dots+ \frac{C_{1 1 2}}{3}x^{-2/3} \psi_2(0)+\ldots\; ,
\eea
and
\be
\psi_3(x) \psi_1(0) = C_{1 2 3} x^{-2/3} \psi_2(0)+\dots + C_{1 3 4} x \psi_4(0)+\dots\; .
\ee
Thus
\be\label{p1p3cf}
L_{-1}\psi_1(x) = \frac{C_{1 1 2}}{3 C_{1 2 3}} \psi_3(x)=\frac{3^{1/4}}{2 \sqrt{\pi}} \psi_3(x)\; ,
\ee
so that $K$ is indeed given by (\ref{Kval1}). Note that  it appears as a ratio of boundary operator product expansion coefficients, rather than the derivative of the weight $h_1 := h_{(1,2)}$ with the respect to the Potts parameter $q$, as in \cite{JC3}.  In fact our result for $K$ also implies that
\be \label{heq}
h_1'(1)=\frac12 (h_2(1) \frac{C_{1 1 2}}{C_{1 2 3}})^2,
\ee
where the evaluations are at $q=1$.

%SECTION
\section{Discussion \label{furdisc}}

In this section, we discuss a few implications of our calculations above, especially the relation (\ref{opid}) (see also (\ref{p1p3cf})).  

The full consequences of (\ref{opid}) remain to be explored.  However, this relation appears to be supported by representation theory, according to which the highest-weightspaces of a Verma module are one-dimensional \cite{AR}, so that any  two primary operators of the same weight must be proportional, as in  (\ref{opid}).  It is also interesting that the integral weights for the $c=0$ primary operators are exactly the Euler pentagonal numbers \cite{RW}.  There are indications that relations similar to  (\ref{opid}) hold for all of them. This suggest the presence of some additional symmetry for conformal field theory with $c=0$.

Next, consider the seventh-order null vector, which again factorizes in two ways:
\be\label{7nullrel}
\mathcal{D}_{3,2} \mathcal{D}_{1,1} = \mathcal{D}_{1,5} \mathcal{D}_{1,2}\; .
\ee
Thus, arguing as above, one finds that $L_{-1}\psi_1$ obeys $\mathcal{D}_{3,2}$ as well as $\mathcal{D}_{1,4}$ null vector conditions.  

Consider now the fusion rules of an arbitrary Kac table operator with $\phi_{1,4}$ and $\phi_{3,2}$.  In general one has 
\bea
\label{14fusion}[\phi_{1,4}] \times [\phi_{r,s}] &=& [\phi_{r,s-3}]+[\phi_{r,s-1}]+[\phi_{r,s+1}]+[\phi_{r,s+3}]\\
\nonumber [\phi_{3,2}] \times [\phi_{r,s}] &=& [\phi_{r-2,s-1}]+[\phi_{r,s-1}]+[\phi_{r+2,s-1}]+[\phi_{r-2,s+1}]\\
\label{32fusion}&&\qquad+[\phi_{r,s+1}]+[\phi_{r+2,s+1}]
\eea
The above then implies that only families present in both of these should be contained in the $L_{-1}\psi_1$ fusion rule.  This leads to 
\be \label{dpsi1fus}
[L_{-1}\psi_1] \times [\phi_{r,s}] = [\phi_{r,s-1}] + [\phi_{r,s+1}] = [\psi_1] \times [\phi_{r,s}]\; .
\ee
Since $\psi_1$ and $L_{-1}\psi_1$ belong to the same conformal family, they should transform among the same conformal families under fusions, in agreement with (\ref{dpsi1fus}).

Thus, our use of $\psi_3$ to obtain the crossing densities augments the $[\psi_1]$ conformal family.  The two additional families present in (\ref{14fusion}) generate crossing configurations that are more complicated than those that can be generated by $\psi_1$ operators alone.  Specifically the inclusion of the $\psi_3$ operator allowed us to make use of the $[\psi_3] \times [\phi_{r,s}] = [\phi_{r,s+3}]$ fusion which gives configurations of the type shown in Figure \ref{config3}[C].

We can also use the actions of the fifth and seventh level null vectors on the identity operator to deduce properties of the stess tensor $T$.  Now
\bea \nonumber
\mathcal{D}_{1,2} \mathbf{1}(z)  &=& (3 L_{-1}^2 - 2 L_{-2})\mathbf{1}(z)\\
&=&-2  T (z).
\eea
Using (\ref{wattsvec}) and (\ref{7nullrel}) then shows that the stess tensor is annihilated by both $\mathcal{D}_{3,1}$ and $\mathcal{D}_{1,5}$. (Note that when $c = 0$, $T$ is a primary operator.)

Further, as argued for $\psi_3 \propto L_{-1}\psi_1$, only the conformal families contained in both $\phi_{3,1}$ and $\phi_{1,5}$ fusions should appear in fusions with the stress tensor, which yields
\be
[T] \times [\phi_{r,s}] = [\phi_{r,s}]\; .
\ee
This is as expected, since the stess tensor generates conformal transformations of conformal families amongst themselves.

We hope to explore, elsewhere, the consequences of these remarks, including the ``overlap'' of $T$ and $\psi_4:=\phi_{1,5}$ in analogy with the result for $L_{-1}\psi_1$ and $\psi_3$.

%SECTION
 \section{Acknowledgments}
We thank A. Rocha for useful conversations.

This work was supported in part by the National Science Foundation Grants Nos.  DMR-0536927 (PK) and  DMS-0553487 (RMZ) .

%SECTION

\end{document}